\documentclass[twocolumn]{emulateapj}
\usepackage{apjfonts} 
\usepackage{amsmath}
\usepackage{natbib}
\usepackage{graphicx}
\usepackage{enumerate}

\def\s4g{{S$^4$G}}

\def\ser{S{\'e}rsic }

\usepackage{color}
\definecolor{red}     {rgb}{1.00, 0.35, 0.05}
\definecolor{darkred}     {rgb}{0.70, 0.05, 0.05}
\definecolor{lightblue}    {rgb}{0.05, 0.35, 1.00}
\definecolor{blue}          {rgb}{0.15, 0.15, 1.00}
\definecolor{darkblue}    {rgb}{0.05, 0.05, 0.70}
\definecolor{green}        {rgb}{0.05, 0.65, 0.15}
\definecolor{darkgreen}  {rgb}{0.10, 0.55, 0.15}
\definecolor{magenta}  {rgb}{1.00, 0.00, 1.00}
\definecolor{purple}  {rgb}{1.00, 0.30, 0.30}

\shorttitle{Cosmic Evolution of Bars}
\shortauthors{Kim et al.}

\begin{document}
\title{Cosmic Evolution of Barred Galaxies up to $\MakeLowercase{z}$ $\sim$ 0.84}
\author{
Taehyun Kim\altaffilmark{1},
E. Athanassoula\altaffilmark{2},
Kartik Sheth\altaffilmark{3},
Albert Bosma\altaffilmark{2},
Myeong-Gu Park\altaffilmark{1},
Yun Hee Lee\altaffilmark{4},
Hong Bae Ann \altaffilmark{5}
}
\email{mgp@knu.ac.kr}

\altaffiltext{1}{Department of Astronomy and Atmospheric Sciences, Kyungpook National University, Daegu, 41566, Korea}
\altaffiltext{2}{Aix Marseille Univ, CNRS, CNES, LAM, Marseille, France}
\altaffiltext{3}{Mary W Jackson NASA Headquarters, 300 E Street SW, Washington DC 20546, USA}
\altaffiltext{4}{Korea Astronomy and Space Science Institute, Daejeon, 34055, Korea}
\altaffiltext{5}{Department of Earth Science Education, Pusan National University, Busan, 46241, Korea}

\begin{abstract}
We explore the cosmic evolution of the bar length, strength, and light deficit around the bar for 379 barred galaxies at $0.2 < z \leq 0.835$ using F814W images from the COSMOS survey. Our sample covers galaxies with stellar mass $10.0 \leq \log(M_{\ast}/M_{\odot}) \leq 11.4$ and various Hubble types. The bar length is strongly related to the galaxy mass, the disk scale length ($h$), $R_{50}$, and $R_{90}$, where the last two are the radii containing 50 and 90\% of total stellar mass, respectively.
Bar length remains almost constant, suggesting little or no evolution in bar length over the last $7$ Gyrs. The normalized bar lengths ($R_{\rm bar}/h$, $R_{\rm bar}/R_{50}$, and $R_{\rm bar}/R_{90}$) do not show any clear cosmic evolution.
Also, the bar strength ($A_2$ and $Q_b$) and the light deficit around the bar reveal little or no cosmic evolution.
The constancy of the normalized bar lengths over cosmic time implies that
the evolution of bars and of disks is strongly linked over all times.
We discuss our results in the framework of predictions from numerical 
simulations. We conclude there is no strong disagreement between our results and up-to-date simulations.
\end{abstract}

\section{Introduction}
Bars have a dominant influence on the structuring of disk galaxies, and understanding their evolution is a major topic in galaxy dynamics. They are relatively prominent features in about a third of present day disk galaxies, and weak bars exist in yet another third (e.g. \citealt{eskridge_00, menendez_delmestre_07}).
However, the frequency of bars is found to evolve with redshift.
Some earlier studies have found that bar fractions remain constant out to $z \sim 1$ (\citealt{elmegreen_04_fbar, jogee_04}). However, studies with the dataset covering a larger mass range have found that the bar fraction decreases with redshift (\citealt{abraham_99, sheth_08, cameron_10, melvin_14}). 
The bar fraction is $65\%$ at $z \sim 0.2$ and decreases down to $20\%$ at $z \sim 0.84$ (\citealt{sheth_08}).
The bar fraction at $0.7 \leq z < 1.5$ is reported to be constant at the $10\%$ level, and the highest redshift of an observed barred galaxy found to date is $z=1.5$ (\citealt{simmons_14}).
\citet{kraljic_12}, using simulations, predict that the bar fraction increases with time at $0 < z < 1$, but bars are almost absent at $z > 1$ . A recent study on the IllustrisTNG galaxies (\citealt{zhao_20}) finds that the bar fraction decreases with redshift at $0 < z < 1$ if the progenitors of massive galaxies at $z = 0$ are traced. However, if all disk galaxies are selected with a constant mass cut $(\log(M/M_{\ast})>10.6)$ during $z=0 \sim 1$, the bar fraction remains at $\sim 60\%$.

Bars at $z > 1$ are rare and could be short-lived, as would be expected due to either frequent violent mergers (e.g., \citealt{conselice_03, kartaltepe_10, lotz_11}), dynamically hot disks (e.g., \citealt{kassin_12, kraljic_12}), or the violent disk instability which produces clumpy disks (e.g., \citealt{forster_schreiber_06, martig_12, kraljic_12, conselice_14}). These mechanisms destroy bars or prevent disks from forming them. However, as dynamically cold disks start to settle down at $z \sim 1$, bars are preferentially found in rotation-dominated galaxies (e.g., \citealt{sheth_12}), and most of the bars formed at $z<1$ persist to be long-lived (e.g., \citealt{kraljic_12}).  
Thus, numerical simulations suggest that present-day disk galaxies experience two evolutionary phases: early ``violent'' phase at $z>1$ and late ``secular'' phase at $0<z<1$ (\citealt{martig_12, kraljic_12}). 
The transition between these two phases could provide an important epoch, which is would be linked to the bar formation epoch (\citealt{kraljic_12}).
More massive and redder galaxies form their bars first (\citealt{sheth_08}), and cosmological simulations also find the same trend (\citealt{kraljic_12}). 
Thus, the bar formation epoch may vary with the galaxy mass to some extent. We expect that disks become cold enough to form bars near $z \sim 1$.
%It might be interesting to examine the evolution of the bar length and its strength as a function of redshift for galaxies at the secular phase $z<1$.
Once bars formed, they are expected to evolve in length and strength in the secular phase. It is, therefore, interesting to examine the bar length and strength as a function of redshift for galaxies at the secular phase.

\citet{perez_12} examined 44 barred galaxies with an outer ring at $0<z<0.8$ using data from the SDSS (18 galaxies) and the COSMOS (26 galaxies) dataset. 
They measured the outer ring radius ($R_{\rm ring}$) and bar radius ($R_{\rm bar}$). The ratio of $R_{\rm ring}$ and $R_{\rm bar}$ is related to the bar corotation radius to bar radius, $\cal{ R}$=$R_{\rm CR}/R_{\rm bar}$. Using $\cal{ R}$ as an indicator of the bar pattern speed, they explored the evolution of the bar pattern speed with redshift and found that $R_{\rm CR}$, $R_{\rm bar}$, and the bar pattern speed have not evolved with redshift in the last 7 Gyrs.

Before we explore the bar length evolution, we should keep in mind that the bar length is related to various galaxy properties.
Bars in early-type disk galaxies are typically longer than those in late-type disk galaxies (e.g., \citealt{elmegreen_85, ann_87, martin_95, erwin_05_bar, menendez_delmestre_07, laurikainen_07, hoyle_11, diaz_garcia_16}). However, bars at the very early types (e.g., S0s, T$\leq 0 $) tend to become shorter (\citealt{erwin_05_bar, laurikainen_07, diaz_garcia_16}), whereas those of very late types  (e.g., $6<T<8$) are slightly longer (\citealt{diaz_garcia_16, lee_20}).
In early work, the bar length has been reported to scale with the galaxy luminosity (\citealt{kormendy_79}) and disk scale length (\citealt{ann_87}). It is also strongly related to the galaxy stellar mass (\citealt{erwin_19}, Figure 20 of \citealt{diaz_garcia_16}). \citet{erwin_19} found that for massive galaxies ($\log (M_{\ast}/M_{\sun}) > 10.1$), the bar length is a strong function of the galaxy stellar mass, whereas for less massive galaxies, $\log (M_{\ast}/M_{\sun}) \leq 10.1$, it is almost independent of the galaxy stellar mass. They also found that the bar length strongly correlates with the disk scale length and the effective radius of the galaxy. 

% For Max mu, A2, Qb evolution
Due to their non-axisymmetry, bars can
cause the gas to flow inwards (e.g., \citealt{athanassoula_92b, regan_99, sheth_00}), 
%and this brings the 
thus enhancing star formation in the central regions of their host galaxies (\citealt{sersic_67, hawarden_86, ho_97, sheth_05, ellison_11, coelho_11}). As a result, barred galaxies have various stellar structures, such as nuclear rings, nuclear disks, inner rings, outer rings, and boxy peanut bulges (e.g., \citealt{combes_85, athanassoula_92b, sellwood_93, kormendy_04, knapen_02, comeron_10, kim_w_12a, erwin_13, buta_15, emsellem_15}, and references therein).
These are dense regions in terms of stellar density. However, bars also develop sparse regions around themselves, usually producing a "$\Theta$" shaped region. These sparse regions are thought to be produced by the bar-driven secular evolution in a way that the bar sweeps out materials around it. Such sparse regions are called by different names, such as ``light deficit''(\citealt{kim_16}), ``star formation desert'' (\citealt{james_15, james_16}) and ``dark spacer'' (\citealt{buta_17b}).
In this respect, some barred galaxies have been found to be accompanied by very faint or almost no disk around the bar (\citealt{gadotti_03, gadotti_08, lokas_21}). Using $H\alpha$ imaging, \citet{james_09} found that star formation is suppressed in the immediate vicinity of the bar (\citealt{james_16, james_18}). The bar effectively transports gas into the central region once it is formed. Therefore, the star formation is truncated in the star formation desert. \citet{james_16, james_18} measured the age of stellar populations and put constraints on the bar formation epoch. With zoom-in cosmological simulations, \citet{donohoe_keyes_19} found a truncated star formation and a lack of young stellar populations. However, they stressed that the interpretation can be complicated due to the radial migration of stars. 
\citet{kim_16} found that the light deficit is more pronounced among galaxies with a longer and stronger bar. 
%This suggests that more evolved bars should show a more pronounced light deficit. 
Later, \citet{buta_17b} found that these sparse regions occur not only inside the inner ring, but also between the bar and the outer ring. He also found that the degree of the light deficit is a strong function of the bar strength, $A_2$. Therefore, the bar driven secular evolution is recorded in the light deficit and how it changes with redshift is worthy of investigation.

%\abcom{start with known stuff.}
%This wealth of observational data on bars and their
%evolution was followed by a large number of corresponding theoretical work, mainly  
Early numerical simulations showed that bars can
form naturally in 
pre-existing, axisymmetrically stable, but otherwise unstable
galactic disks (\citealt{miller_70, hohl_71, ostriker_73, sellwood_80}).
Later, pure N-body simulations have found that as a barred galaxy evolves, the bar becomes longer and stronger with time (e.g., \citealt{combes_93, debattista_00, athanassoula_03} (hereafter A03); \citealt{martinez_valpuesta_06}).
With the advances of computer technology, it became possible to consider more realistic cases, including gas and its physics (star formation, feedback and cooling), halo triaxiality and spin, or the effect of companions and interactions (\citealt{berentzen_06, machado_atha_10, athanassoula_13, lokas_14, collier_18, seo_19}, etc.)

The formation and evolution of bars can best be described by distinguishing several phases (\citealt{athanassoula_13, athanassoula_13_book} for a review [hereafter A13]). 
The bar forms from an instability 
of the disk during the first, short phase, which is often  
called the bar formation time. The next one or two phases, also have a short duration and are linked to the vertical instability and the boxy/Peanut/X shape. The last phase covers a much longer time range, known as secular evolution time, during which the bar strength
and length increase slowly but steadily. It is this longer period of time and
the corresponding evolution that will concern us mainly in this paper.

The type of simulations discussed in the previous paragraph can have very high resolution, but have a disadvantage, namely that their initial conditions are idealized, since they assume that the disc has formed and reached equilibrium first, and the formation and evolution of the bar follows. Nevertheless, these simulations and particularly the later, more realistic ones, allowed great progress to be made in our understanding of how and why bars form and evolve, and how they affect the evolution of other components such as the disc and the bulges. Many cases have shown good agreement with observations.

A more realistic situation can occur if the proto-galaxies include no disk, just a dark matter halo and a hot gaseous circum-galactic medium. \citet{athanassoula_16b} followed the evolution of this gas, either in galaxies in isolation or after an interaction and merging of two proto-galaxies. These show that a disk forms, with properties as those observed, i.e. its surface density has a break, separating two exponential parts. A bar forms at the same time as the disk, without waiting for the latter to grow first. The morphology of both the disk and the bar agrees well with observations.

An alternative is to use cosmological zoom re-simulations. Their linear resolution is still far from that of e.g. Athanassoula et al. (2016), which is 25 or 50 pc, but is of the order of what was used in simulations with idealized initial conditions some 10 to 20 years ago, i.e. sufficient for studying a number of aspects of bar formation and evolution.  We will thus use them, together with simulations with idealized initial conditions, to better understand the implications of our results.

This study aims to investigate the evolution of the bar length and strength over cosmic times and of the light deficit produced by the bar which have not been systematically explored yet for all types of barred galaxies.  
The remainder of this paper is organized as follows: \S 2  describes our data and how we selected the galaxy sample; \S 3 gives a brief overview of data analysis; \S 4 presents the results on the cosmic evolution of barred galaxies; \S 5 discusses our results; and \S 6 summarizes and concludes this study.

\section{Data and Sample Selection}
\label{sec:data}
We used the data from the Cosmic Evolution Survey (COSMOS; \citealt{scoville_07_overview, Koekemoer_07}), which surveyed 2 $\rm{deg^2}$ of the sky with the HST ACS. We made use of the F814W images.
Our data includes barred galaxies of both $T \ge 2$ (late-type disk galaxies) and $T < 2$ (early-type disk galaxies), where T-type classification is drawn from \citet{capak_07}.
First we selected galaxies satisfying the requirements of both visual 
and analytical classification for barred galaxies using the ellipticity and the position angle (hereafter PA) criteria of \citet{sheth_08} who explored disk galaxies with a safe cut of $T \ge 2$. In order to supplement barred lenticular galaxies (i.e., $T<2$), we followed the same steps of the sample selection as performed in \citet{sheth_08}. The criterion is plotted as a dashed line in Figure 1(a). Detailed sample selections are referred to \citet{sheth_08}. For $T<2$ galaxies, we selected galaxies that two of the authors (T. K. and H. B. A.) agreed to visually classify as barred. 
Briefly, we selected all disk galaxies that are brighter than $L^{*}_{V}$, which is the
empirically determined luminosity evolution from \citet{capak_04_phd} which satisfies $M_V$ = $-21.7$ mag at $z=0.9$. This condition is set to trace galaxies from the same portion of the galaxy luminosity distribution at all redshifts we explore. Details on the sample selection are 
given in \citet{sheth_08}. The redshifts of the galaxies are drawn from the COSMOS2015 Catalog (Laigle et al., 2015) which provides photometric redshifts.
We present the $M_{V}$ and the mass distribution of barred galaxies that we finally chose in Figure 1. The mass of the barred galaxies ranges from $\log (M_\star/M_\sun)=$ 10.0 to 11.4. We made use of the stellar mass estimates from \citet{mobasher_07} throughout this paper.
The mass distribution of the galaxies in Figure 1(c) shows that lower-z galaxies ($0.2 < z 
\leq 0.5$) are skewed to a lower mass, while higher-z galaxies ($0.5 < z \leq 0.835$) are normally distributed. Overall, two redshift groups span the same mass range.
Figure 1(d) shows that the mass distribution depends on the T-type of galaxies. Early-type disk galaxies dominate more massive bins, while late-type disk galaxies dominate less massive bins.

Bars are composed of relatively old stellar populations, and can be best traced by near infrared that is moderately free of extinctions.
Meanwhile, bars are almost invisible at UV (\citealt{sheth_03}), where the light traces young stellar populations that actively form stars. Thus, it is important to limit the redshift range of the samples. To that aim, we select galaxies within $0.2 < z \leq 0.835$ considering the band shifting effect (\citealt{sheth_03}), as the wavelength of F814W becomes almost the B-band at $z = 0.835$.

As we will state in the next section, we also estimate bar length and decompose galaxies into bar, disk, and bulge, if any. Some barred galaxies were removed from the sample due to unreliable results in two-dimensional decompositions. Consequently, we ended up having 379 barred galaxies in our sample.

%--------------------------------------------------------------------------------------
% Figure:1
%--------------------------------------------------------------------------------------
\begin{figure*}
%\begin{figure}
\begin{center}
\includegraphics[width=\textwidth]{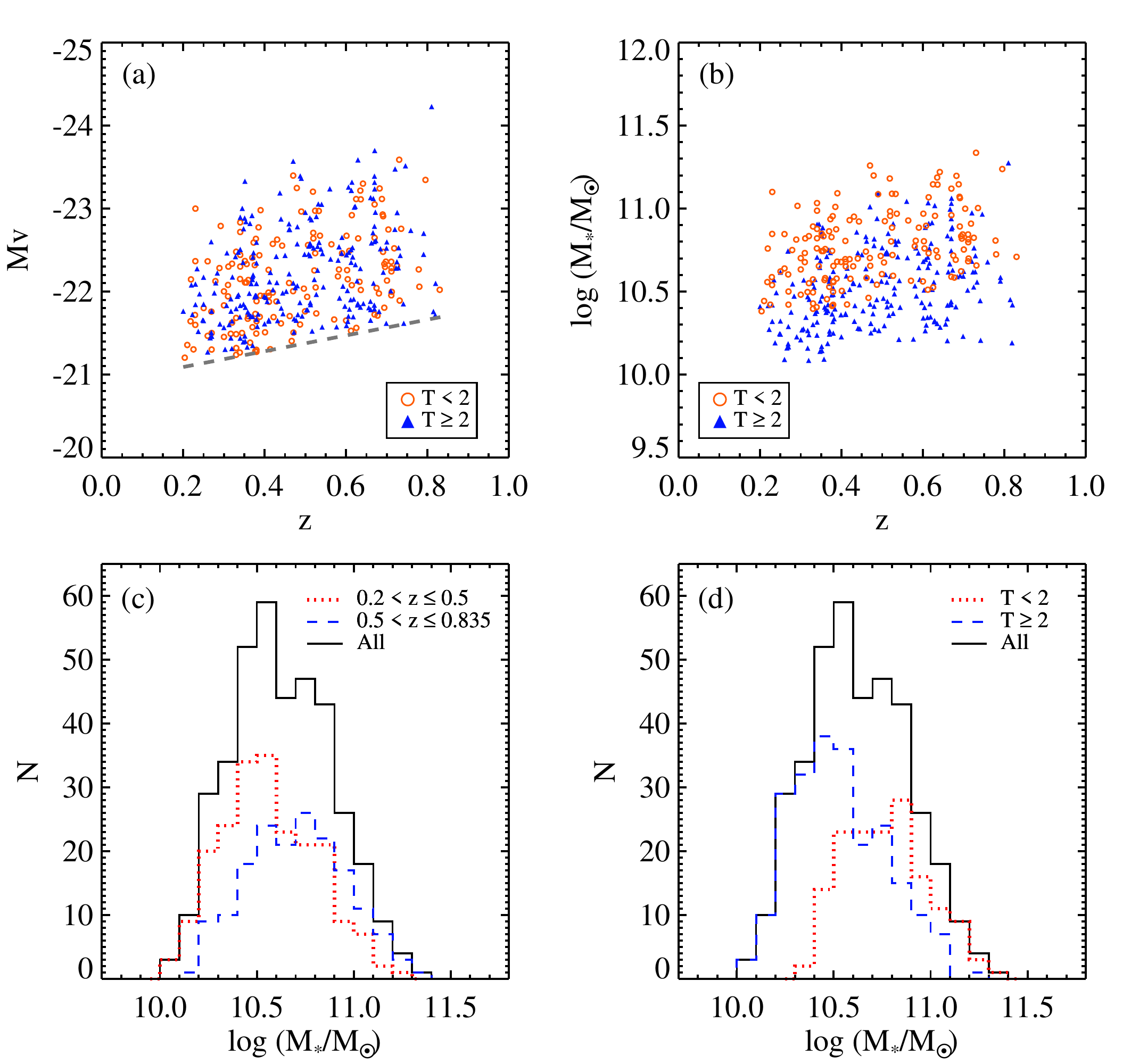}
        \caption{(a) Redshift and absolute V band magnitude ($M_{V}$) of the sample galaxies from \citet{capak_07}. The dashed line denotes $L^{*}_{V}$ which is empirically determined luminosity evolution from \citet{capak_04_phd}.
        (b) Redshift and stellar mass of the sample galaxies: Galaxies with T $<2$ are plotted in orange circles, while galaxies with T $\geq$ 2 are plotted in blue triangles.
        (c) Mass distribution of the sample galaxies: galaxies at $0.2 < z \leq 0.5$ are denoted by red dotted line, while the galaxies at $0.5 < z \leq 0.835$ by blue dashed line. 
        (d) Mass distribution of galaxies: T $<2$ (early-type) barred disk galaxies are in red dotted line and  T $\geq$ 2 (late-type) barred disk galaxies are in blue dashed line.
        }
\label{fig:mass_z}
\end{center}
\end{figure*}
%\end{figure}
%--------------------------------------------------------------------------------------

\section{Data analysis} 
\label{sec:data_analysis}
\subsection{Bar length estimation}
\label{subsec:barlength}
%Fig: Bar length estimation
The bar length has been measured using various methods, including visual estimation (e.g., \citealt{kormendy_79, martin_95, hoyle_11, herrera_endoqui_15}), 
fitting isophotes of galaxy and use ellipticity, PA profiles (e.g., \citealt{wozniak_91, wozniak_95, knapen_00, sheth_03, erwin_05_bar, menendez_delmestre_07, munoz_mateos_13}),
Fourier analysis (e.g., \citealt{elmegreen_85, ohta_90, aguerri_00b, athanassoula_02a, laurikainen_05, diaz_garcia_16, lee_19}),
two-dimensional decompositions (e.g., \citealt{ann_87, prieto_01, peng_02, laurikainen_05, gadotti_08, kim_14, salo_15}), and
light profile along the major axis of the bar (e.g., \citealt{elmegreen_85, seigar_98a, chapelon_99}).
By testing various bar length measuring methods, \citet{athanassoula_02a} presented pros and cons for each method. \citet{michel_dansac_06} showed that one has to select the right bar length estimator depending on the application as each method does not probe the same dynamical characteristic radii of bars, i.e., bar resonance radii such as ultraharmonic radius or corotation radius.
As we plan to compare bar length systematically but not to further study bar resonance radius nor their ratio in this study, we fit isophotes to the galaxy and inspect the change in the ellipticity and PA profile to determine the bar length. We estimate the radius where the ellipticity profile reaches its maximum ($R_{\rm emax}$). It has been shown that $R_{\rm emax}$ tends to underestimate the bar length (e.g., \citealt{wozniak_95, athanassoula_02a, laurikainen_02_2mass, erwin_03}). Thus, we define $R_{\rm emax}$ as the lower limit of the bar radius. For the upper limit of the bar radius, we consider following three radii and select their minimum as the upper limit. First, we estimate the radius where the ellipticity reaches its local minimum ($R_{\rm emin}$) after $R_{\rm emax}$.  Second, we estimate the radius where the elliptiticy decrease by 0.1 from the maximum ellipticity ($R_{\Delta e0.1}$). Third, we measure the radius $R_{\Delta PA10}$ at which the PA change by 10 degree from the PA at $R_{\rm emax}$. We then estimate the radius of the bar by obtaining the mean of the lower and upper limits of the bar radius. 

From z $=0.2$ to $0.835$, the physical resolution of the HST COSMOS images changes from $0.3$ to $0.72$ kpc (i.e. $\sim 2.4$ times).  In order to test the effect of resolution, we took 44 galaxies at $0.2<z<0.3$ and artificially binned the data by $2 \times 2$ and $3 \times 3$. We found that the bar length decreases by $1.5\%$ and $4.3\%$ when we bin the data by $2 \times 2$ and $3\times 3$, respectively. Therefore, the effect of resolution should be taken into account.

Note that the bar length is difficult to measure accurately. Furthermore, various methods can come with different answers. Thus, our measurements may, by necessity, include some amount of random error and/or some bias. This is a general caveat to all studies including bar lengths.

% Bar length estimation 
% R_lower= R_emax
% R_upper= min(R_emin, R_pa10, R_e01)
% R_bar = mean( R_lower, R_upper)
%The median bar ellipticity is XXX

\subsection{Bar length at different wavelengths}
%Fig: SINGS sample
The data contain galaxies of a wide range of redshift from 0.2 to 0.835. The restframe mean wavelength of F814W varies from 6783{\AA} ($\sim \rm{R-band}$) at $z=0.2$ to 4436{\AA} ($\sim \rm{B-band}$) at $z=0.835$. Therefore, it is necessary to check whether the bar length changes with respect to wavelength. We took $B,V,R$ and $I$ band images of 11 barred galaxies from the SINGS Survey (\citealt{Kennicutt_03}), and measured the bar length and find that the bar lengths agree within, on average, 5$\%$ in these four passbands. However, for 2 out of 11 galaxies (NGC 1512 and NGC 7552), the bar length measured in the $B$ or $I$ band stands out due to the bright star forming knots at the end of the bar. This leads the bar length to change up to $\sim 15\%$. We will take this into account for later interpretation. 

While we obtained 5$\%$ differences on average, using more statistically complete samples \citet{gadotti_11} found that the effect of wavelength change in bar size is negligible. \citet{gadotti_11} performed two dimensional decompositions on $\sim$ 300 barred galaxies from SDSS, and found that the change in bar size is less than 1$\%$ among $\rm g-,\rm r-,\rm i-$band.

\subsection{Two-dimensional decomposition using {\sc GALFIT}}
We perform a two-dimensional (2D) multi-component decomposition using {\sc GALFIT} (\citealt{peng_02, peng_10}). We decompose galaxies into bulge (if any), disk and bar. We fit bulges with the \ser profile (\citealt{sersic_63}) defined as
\begin{equation}
%\begin{split}
\Sigma(r) = \Sigma_{e} \exp \left[-\kappa \left( \left(\frac{r}{r_e}\right)^{1/n} - 1\right) \right],
\label{eq:eq_sersic}
%\end{split}
\end{equation}
where $r_e$ is half light radius and $\Sigma_e$ is the surface brightness at $r_e$ (\citealt{caon_93}). The \ser index, $n$, describes how the light is distributed in the central region, 
and $\kappa$ is a variable related to $n$.
When $n$ is large, light is centrally concentrated and has a steep profile in the inner part and an extended profile in the outer part. If $n$ is small, however, the light has a shallow inner profile and an abruptly truncated outer profile. 

Disk components are fit with the Exponential profile defined as
\begin{equation}
%\begin{split}
\Sigma(r) = \Sigma_{0} \exp \left( - \frac{r}{h} \right),
\label{eq:eq_exponential}
%\end{split}
\end{equation}
where $h$ is the disk scale length. The Exponential profile is a special case of the \ser profile when the \ser index $n$ is fixed to 1. 

We adopt a modified Ferrers profile for the bar component, defined as follows:
\begin{equation}
%\begin{split}
\Sigma(r) = \Sigma_{0} \left[ 1- \left(\frac{r}{r_{\rm out}}\right)^{2-\beta} \right]^\alpha.
\label{eq:eq_ferrer}
%\end{split}
\end{equation}
The profile is defined out to $r = r_{\rm out}$, and we choose $r_{\rm out}$ to be the bar length that we estimated using the ellipticity and the PA profiles of the galaxies, as described in Section 3.1. 
The bar lengths from the 2D decompositions tend to yield longer bar lengths than those provided by the isophote analysis and visual estimation, because 2D decomposition tries to find the minimum of the $\chi^2$ from the model fit and observations. %% ADD boxy components, ansae part of the bar etcs.
Therefore, in this study, we adopt the bar length from the ellipticity and PA profile methods.
The parameter $\alpha$ describes how sharply the bar truncates at the end, while $\beta$ characterizes the central slope of the bar radial profile.

The {\sc GALFIT} results depend on the input parameters, especially in multi-component fits. For example, if we provide initial input parameters far from the structural parameters of the galaxy of interest to {\sc GALFIT}, dominant components may supress less prominent components in the first several iterations. Therefore, it is important to provide reliable input parameters to {\sc GALFIT}. To ensure this, we start with a simple one-component fit, and add up components one by one. For the one-component model, 
we fit the galaxy with the Exponential disk model. We again fit the galaxy only with a free {\ser} bulge model. Using these two one-component fit results, we build up inputs for the two-component (bulge + disk) model fit. Finally, we add a bar component into the galaxy model fixing the bar radius estimated from Section 3.1, and find the fit to the galaxy with three components.

\subsection{Bar strength: $Q_b$ and $A_2$}
The bar strength is a parameter that describes how strong the bar is compared to the underlying galaxy components, e.g., disk and bulge.
Various definitions have been introduced, including bar ellipticity, bar-to-total mass ratio, non-axisymmetric torque parameter, and amplitude of the Fourier component (e.g., \citealt{buta_01, athanassoula_13}). 
% How we estimate A2, Qb
Here, we use the bar strengths denoted as $A_2$ and $Q_b$ as follows.

%Assuming the the mass-to-light ratio is constant across the disk\citep{quillen_94}, 
We calculate the gravitational potential assuming that the mass-to-light ratio is constant across the disk \citep{quillen_94, diaz_garcia_16, lee_20}. We then construct the transverse-to-radial force ratio map
to determine $Q_b$, which 
%$Q_b$ 
is the maximum of the tangential-to-radial force ratio calculated in  the polar coordinate system. The details on the bar strength calculations are described in \citet{lee_20}.

We also performed the Fourier decomposition following \citet{athanassoula_13},
\begin{equation}
a_m(R) = \sum_{i}{m_i} \cos(m\theta_{i}),
\end{equation}
\begin{equation}
b_m(R)= \sum_{i}{m_i} \sin(m\theta_{i}),
\end{equation}
where $R$ is the cylindrical radius, $m_i$ and $\theta_i$ are the mass and the azimuthal angle of the $i$-th bin at the radius R.

The maximum amplitude of the $m=2$ component, $A_2$, is defined as 
\begin{equation}
A_2 = \rm{max} \left(\frac{\sqrt{a_2^2+b_2^2}}{a_0}\right).
\end{equation}

% Resolution Effect
%From z $=0.2$ to $0.835$, the physical resolution of the HST COSMOS images changes from $0.3$ to $0.72$ kpc (i.e. $\sim 2.4$ times).  In order to test the effect of resolution, we took 44 galaxies at $0.2<z<0.3$ and artificially binned the data by $2 \times 2$ and $3 \times 3$. 
As we have estimated the resolution effect in estimating bar length in Section 3.1, we also estimated the effect of resolution in bar strength. We find that $A_2$ decreased by $0.9\%$ and $6.7\%$ when the data are binned by $2\times2$ and $3\times 3$, respectively. The value of $Q_b$ increased by $6.1\%$ and $5.8\%$ when we bin the data by $2 \times 2$ and $3\times 3$, respectively.

\section{Results}

%--------------------------------------------------------------------------------------
% Figures 2
%--------------------------------------------------------------------------------------
\begin{figure*}
\begin{center}
%\hspace*{-7cm}\vspace{1.0cm}
%\includegraphics[width=16cm]{test_002.pdf}
%\includegraphics[width=16cm, natwidth=576,natheight=405]{test_002.pdf}
%\includegraphics[width=16cm, bb= 110 30 690 579]{figure_002.pdf}
\includegraphics[width=16cm]{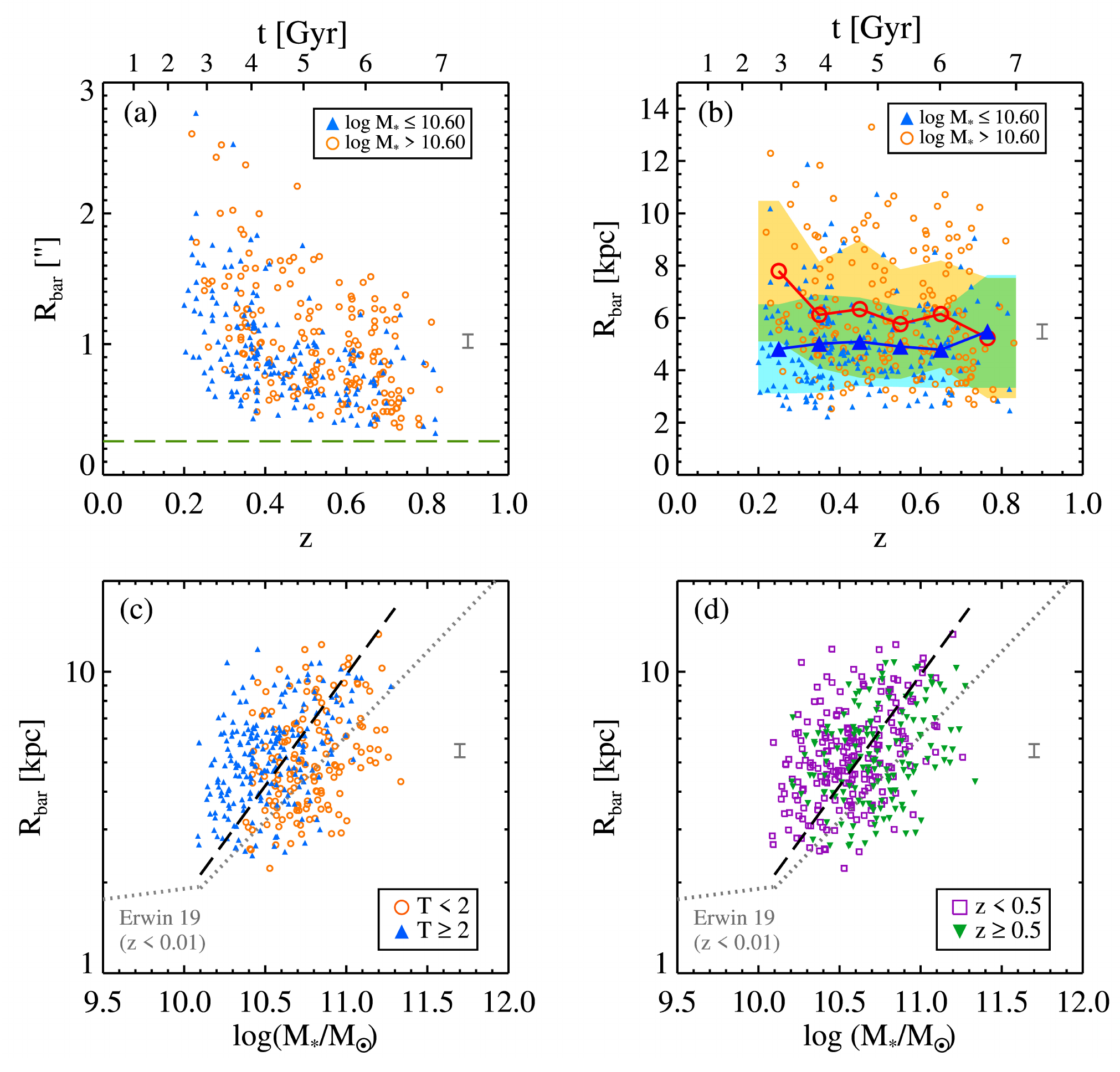}
\vspace{1.0cm}
        \caption{(a) Bar length in arcsec as a function of redshift. 
        The green long dashed line represents our bar detection limit (see text).
        (b) Bar length in kpc as a function of redshift.  
        The green long dashed line traces our bar detection limit. Large symbols which are connected with straight lines represent the mean value at each redshift bin, red circles represent massive galaxies with $\log (M_{\ast}/M_{\odot}) > 10.6$ and blue triangles represents less massive galaxies with $\log (M_{\ast}/M_{\odot}) \leq 10.6$. Orange and blue shaded areas span the standard deviation around the mean value at each redshift bin for massive and less massive galaxies, respectively.
        (c) Bar length as a function of the galaxy mass. Galaxies are color-coded by T-types. The black long dashed line represents the ordinary least square fit to the all data, which is $log(R_{\rm bar})= 0.74 \times log (M_{\ast}/M_{\odot}) -7.11$.
        The gray dotted line shows the locally weighted regression fit for nearby ($D < 40$ Mpc) galaxies from \citet{erwin_19}.
        (d) Same as (c), but galaxies are color-coded by redshifts. 
        All bar lengths are deprojected lengths. We plot the representative error bars of $5\%$ at the mean bar length which takes account of the resolution effect ($<4.3\%$) and wavelength effect ($<5\%$).
       }
\label{fig:rbar_z}
\end{center}
\end{figure*}
%--------------------------------------------------------------------------------------

%-------------------------------------------------------------------------------------
% Figures 3
%--------------------------------------------------------------------------------------
\begin{figure}
\begin{center}
\includegraphics[width=8cm]{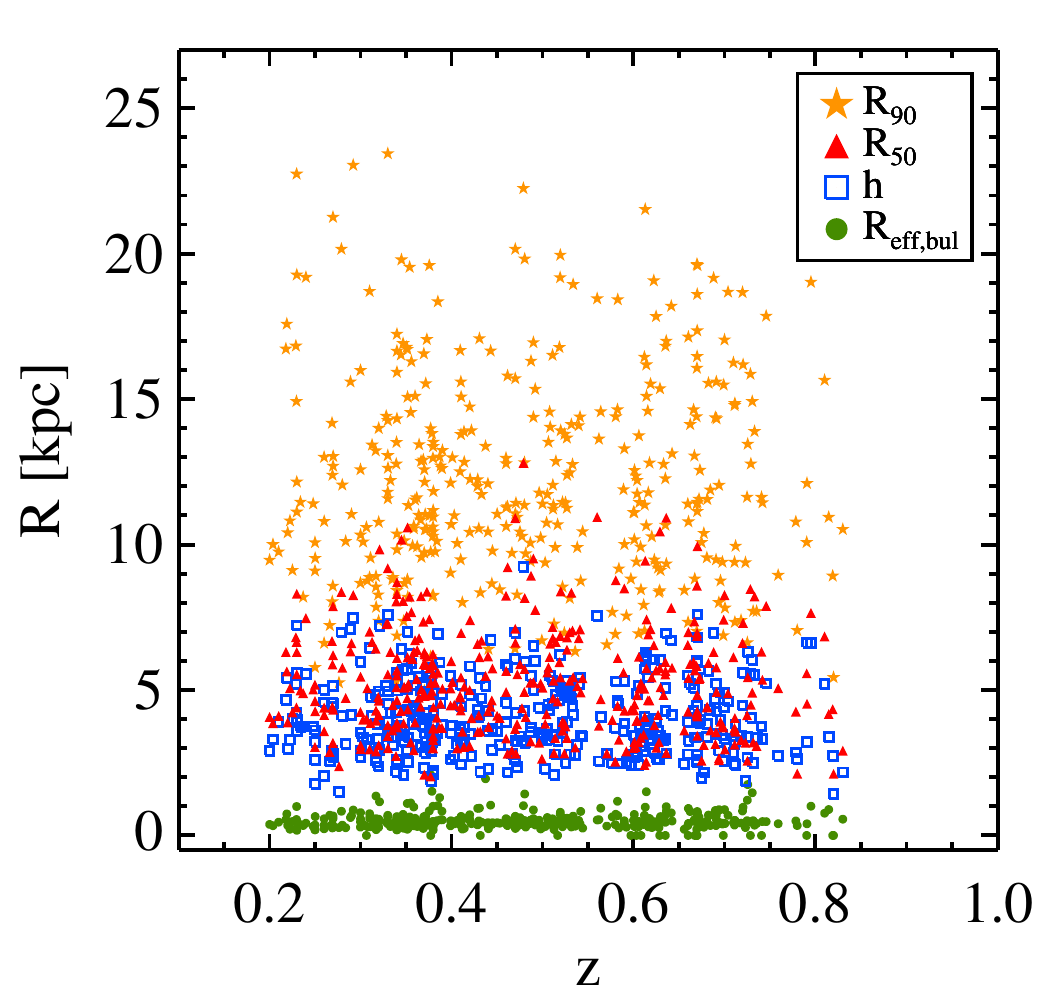}
        \caption{Characteristic radii of galaxies as a function of redshift.
        Effective radius of the bulge component ($R_{\rm eff,\rm bul}$), 
        the disk scale length $(h)$, $R_{50}$, and $R_{90}$ are presented.
        Gray dashed line denotes the linear fit of characterisc radius versus redshift.}
\label{fig:rgal_z}
\end{center}
\end{figure}
%--------------------------------------------------------------------------------------

%--------------------------------------------------------------------------------------
% Figures 4
%--------------------------------------------------------------------------------------
\begin{figure*}
\begin{center}
\includegraphics[width=\textwidth]{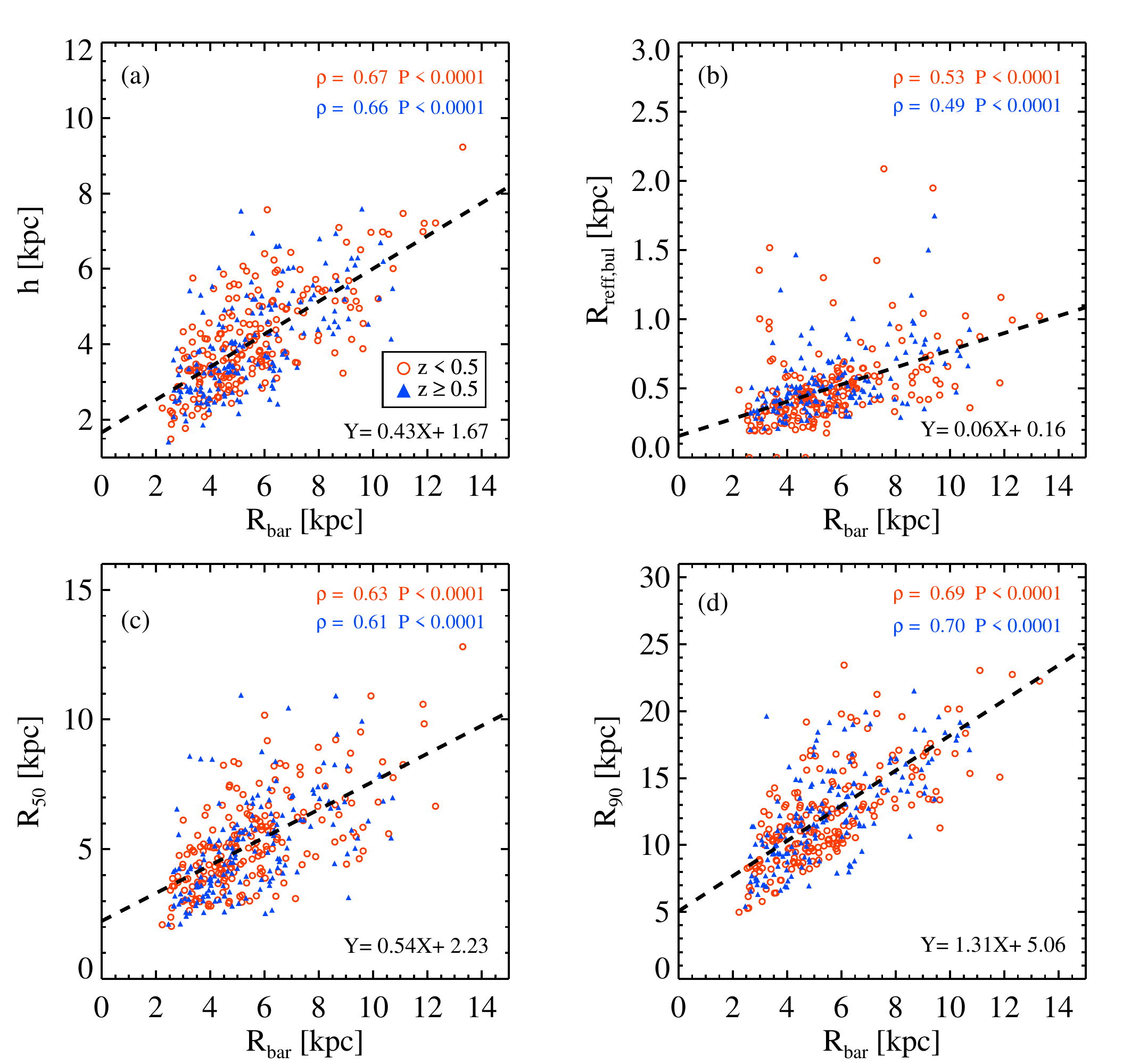}
        \caption{Various characteristic galaxy lengths versus the bar length, all deprojected. 
     Galaxies are color coded by their redshifts. Orange circles represent galaxies at $z<0.5$ and blue triangles represent galaxies at $z \geq 0.5$.
        The Spearman correlation coefficient ($\rho$) and probablity ($P$) are presented in the top right of each panel. Correlation coefficients in the first row (orange color) are results for galaxies at $z<0.5$ and those in the second row (blue color) are results for galaxies at $z \geq 0.5$.
        %The Pearson correlation coefficient ($\rho_p$) is noted in the second line of each panel.
(a) Disk scale length ($h$) as a function of bar length.
(b) Effective radius of the bulge component ($R_{\rm eff,\rm bul}$) as a function of bar length.
(c) $R_{50}$ versus bar length.
(d) $R_{90}$ versus bar length.       
The linear fit to each characteristic radius of galaxies as a function of the bar length is plotted in a dashed line. The coefficients of the linear fit are noted at the bottom right of each panel.}
\label{fig:rbar_galaxy_length}
\end{center}
\end{figure*}
%--------------------------------------------------------------------------------------

 %--------------------------------------------------------------------------------------
% Figures 5
%--------------------------------------------------------------------------------------
\begin{figure*}
\begin{center}
\includegraphics[width=\textwidth]{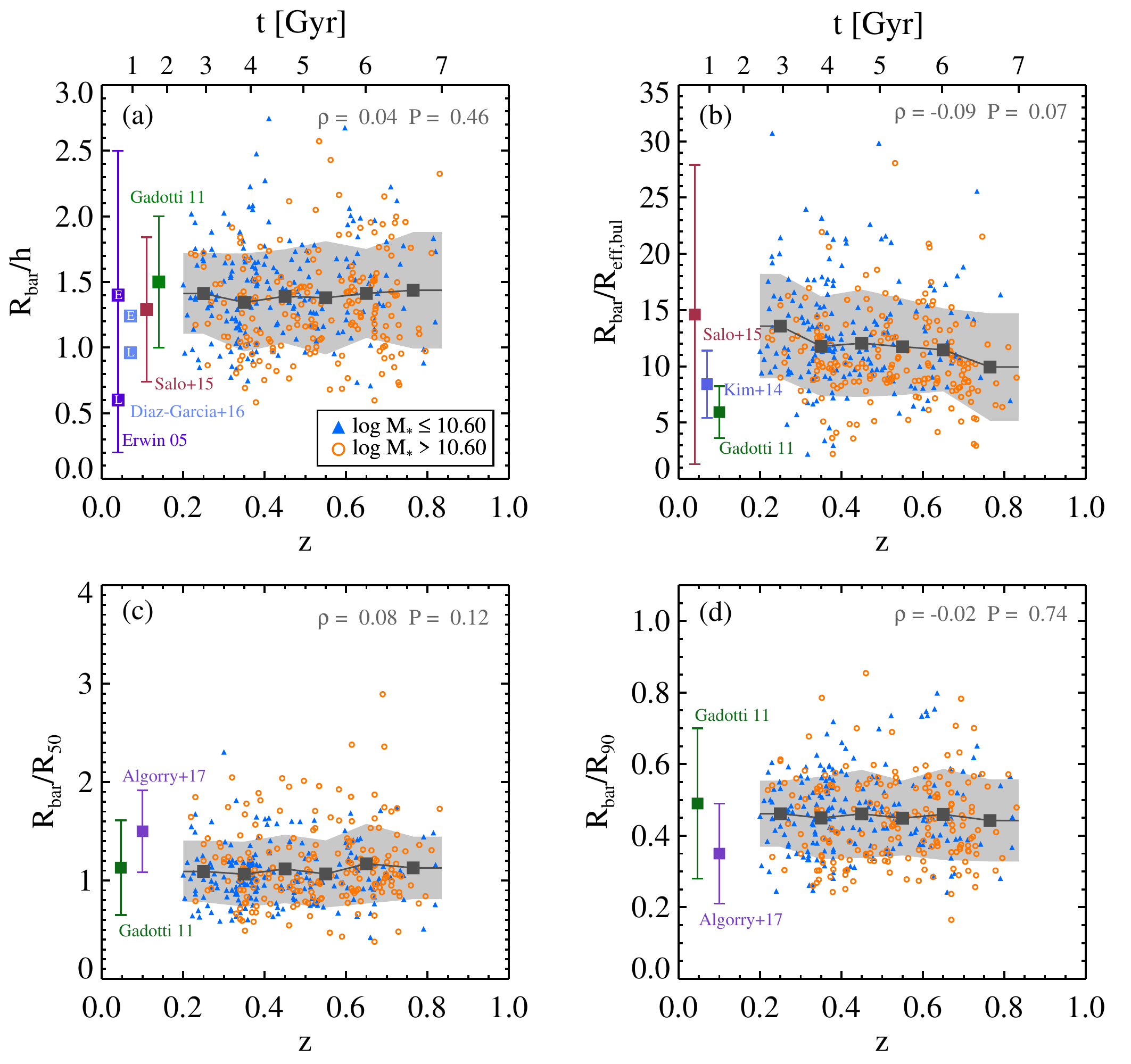}
       \caption{Normalized bar lengths as a function of redshift with  
        (a) $R_{\rm bar}/h$,
        (b) $R_{\rm bar}/R_{\rm eff,\rm bul}$
        (c) $R_{\rm bar}/R_{50}$, and 
        (d) $R_{\rm bar}/R_{90}$. 
        All bar lengths are deprojected. Massive galaxies are denoted as orange open circles, and less massive ones as blue filled triangles. The Spearman correlation coefficient ($\rho$) and statistical significance ($P$) are presented at the upper right of each panel. We binned the data with redshift as performed in Figure 2(b). The black squares denote the mean normalized bar length at each redshift bin. The gray shades represent the standard deviation of each redshift bin. The normalized bar length valies from studies of nearby galaxies are overplotted as colored squares in each panel, for comparison. 
The upper squares marked with ``E'' are for early-type barred galaxies and the lower ones maked with ``L'' are for late type barred galaxies.
\citet{erwin_05_bar} and \citet{diaz_garcia_16} presented the mean $R_{\rm bar}/h$ seperately for early- and for late-type barred galaxies. 
}
\label{fig:galaxy_length_z}
\end{center}
\end{figure*}
%--------------------------------------------------------------------------------------
 %--------------------------------------------------------------------------------------
% Figures 6
%--------------------------------------------------------------------------------------
\begin{figure*}
\begin{center}
\includegraphics[width=\textwidth]{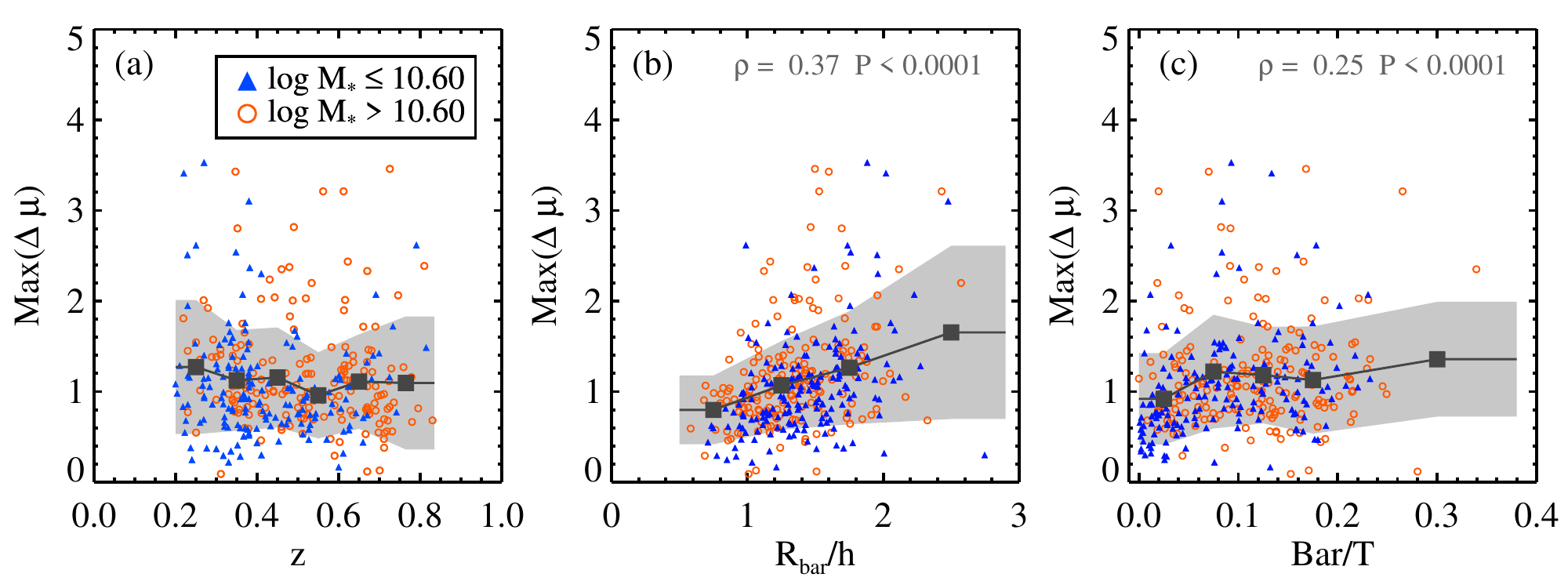}
        \caption{$\rm Max(\Delta \mu$) is the maximum difference between the surface brightness along the bar's major and minor axes. $\rm Max(\Delta \mu$) versus 
        (a) redshift, 
        (b) normalized bar lengths, and
        (c) Bar/T.
The Spearman correlation coefficient ($\rho$) and significance ($P$) are presented at the upper right of each panel. More massive galaxies are plotted in orange open circles and less massive galaxies in blue filled triangles.
        }
\label{fig:mu_max}
\end{center}
\end{figure*}
%--------------------------------------------------------------------------------------
 %--------------------------------------------------------------------------------------
% Figures 7
%--------------------------------------------------------------------------------------
\begin{figure*}
\begin{center}
\includegraphics[width=\textwidth]{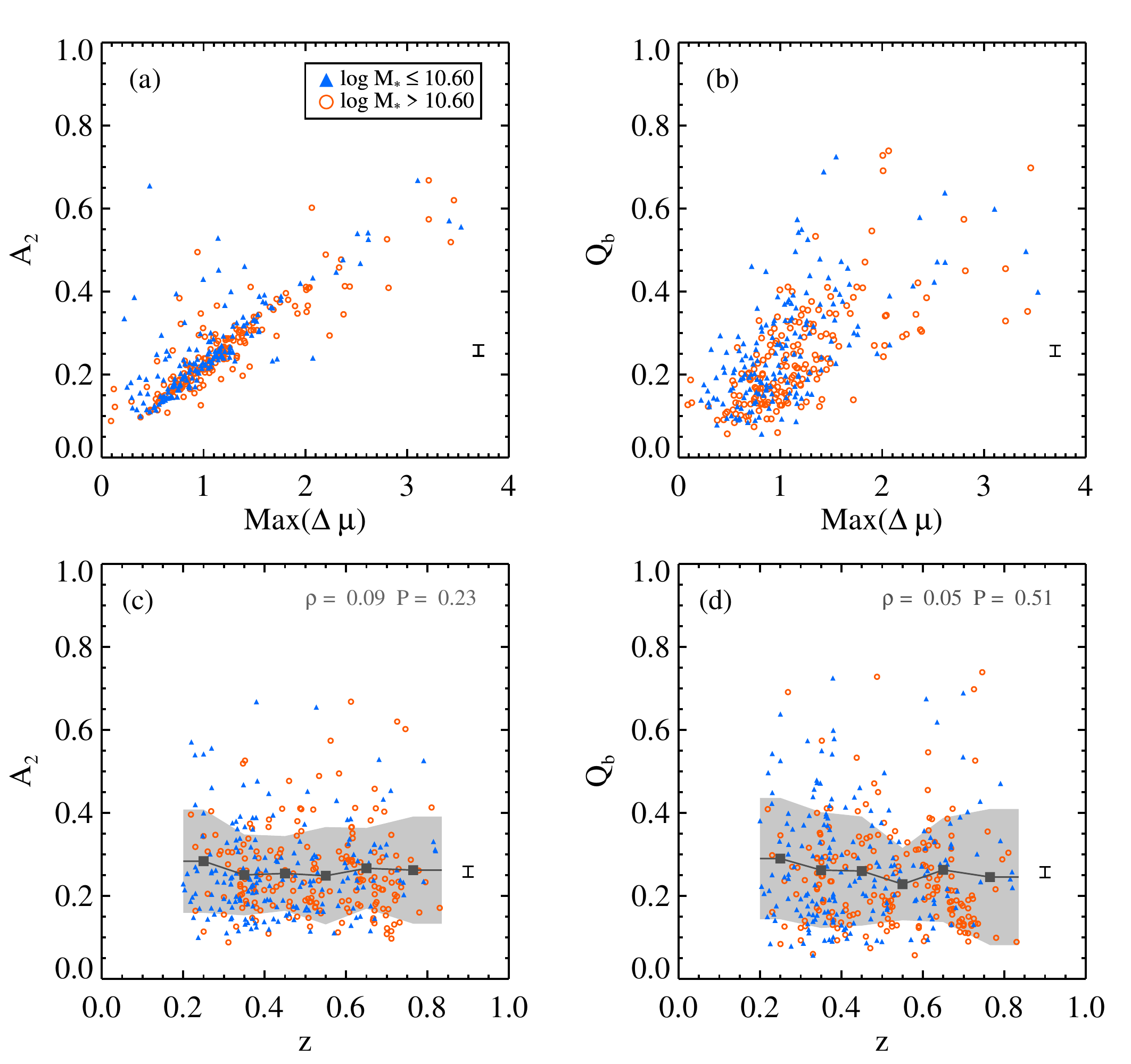}
        \caption{$\rm Max(\Delta \mu$) and the bar strengths (a)$A_2$ and (b) $Q_b$. 
        $A_2$ is the maximum amplitude of the $m=2$ component from the Fourier decomposition and $Q_b$ is the maximum of the tangential to radial force ratio (\citealt{lee_20}). The redshift and  
        (c) $A_2$ and
        (d) $Q_b$. 
        The black squares show the mean value at each redshift bin. The gray shaded area spans the standard deviation around the mean. Massive galaxies are plotted in orange circles, whereas less massive galaxies are in blue triangles.
       We plot the representative error bars of $5\%$ at the mean bar strength which denote the resolution effect.
        }
\label{fig:a2_qb}
\end{center}
\end{figure*}
%--------------------------------------------------------------------------------------
 %--------------------------------------------------------------------------------------
% Figures 8
%--------------------------------------------------------------------------------------
\begin{figure*}
\begin{center}
\includegraphics[width=\textwidth]{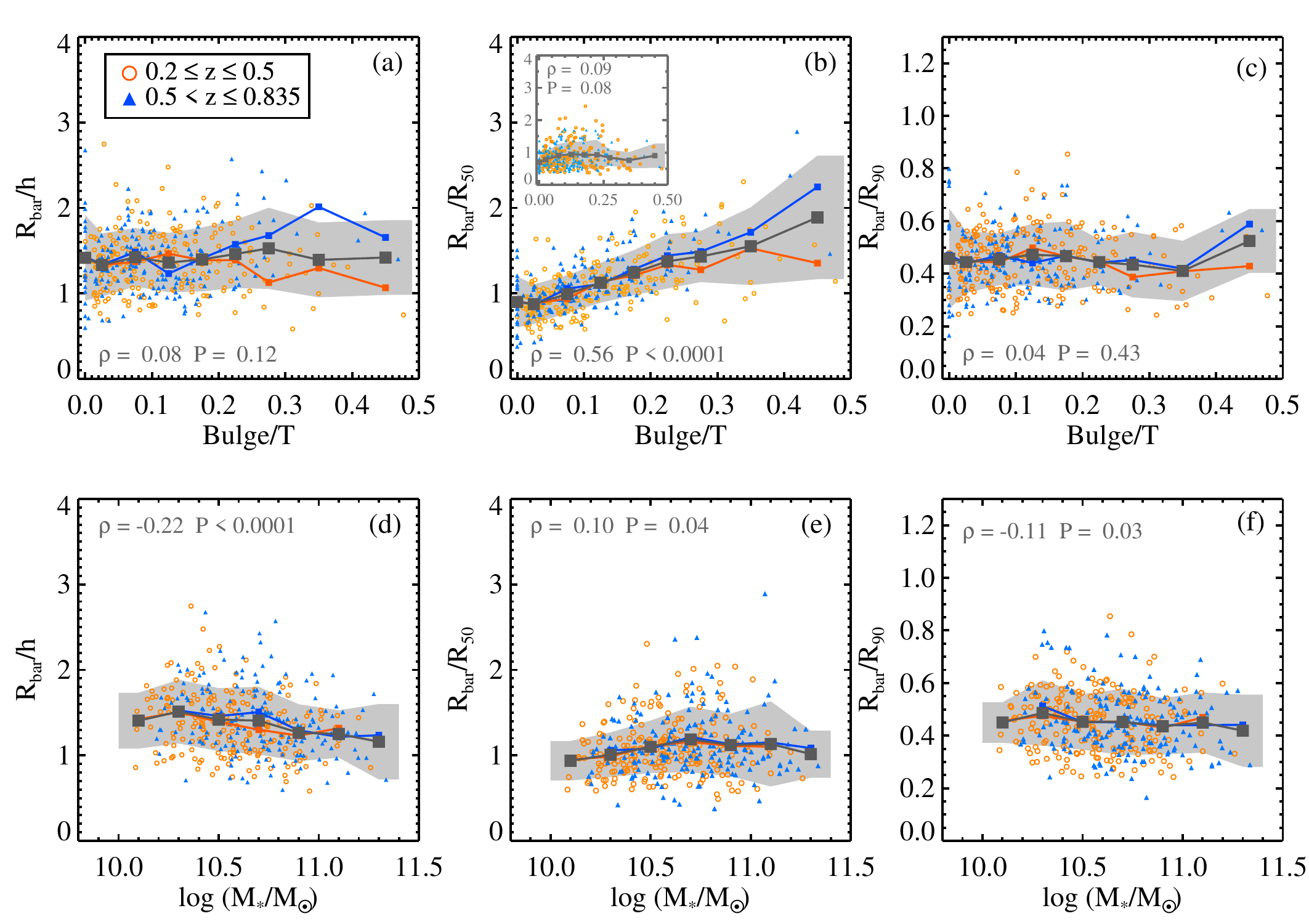}
%\includegraphics[width=16cm, natwidth=765,natheight=540,angle=270,origin=c]{test.pdf}
%\hspace*{-10cm}
%\includegraphics[width=0.8\textwidth, angle=270, origin=c, natwidth=574,natheight=405]{test_009.pdf}
        \caption{Normalized bar lengths as a function of Bulge/T (upper panels) and the galaxy mass (lower panels). The Spearman rank correlation coefficients ($\rho$) and the significance ($P$) are noted in each panel. Low-z galaxies are plotted in orange circles and high-z galaxies in blue triangles. Orange and blue lines represent the mean normalized bar lengths for low-z and high-z galaxies, respectively. The black squares are the mean normalized bar lengths for all samples at each bin and the gray shaded area covers the standard deviation.
        $R_{50}$ is found to be anti-correlated with Bulge/T. Thus, we normalize $R_{50}$ with the linear fit between $R_{50}$ and Bulge/T, and plot the normalized $R_{\rm bar}/R_{\rm 50}$ in the inset of (b).
        }
\end{center}
\end{figure*}
%--------------------------------------------------------------------------------------
\subsection{Cosmic evolution of the bar length}
\label{sec:res1}

%% Fig 2 %%%%%%%%%%%%%%%%%%%%%%%%%%%%%%
We plot the absolute bar length against the redshift in the top panels of Figure 2. The absolute bar length spans $0.3''$ to $2.8''$ (Figure 2a) or $2$ to $13.3$ kpc (Figure 2b). 
The median $\log (M_{\ast}/M_{\odot})$ of our sample galaxies is 10.6. We divide our sample into two mass groups. 
Massive galaxies with $\log (M_{\ast}/M_{\odot}) > $ 10.6 are plotted in orange circles while less massive galaxies with $\log (M_{\ast}/M_{\odot}) \leq$ 10.6 are plotted in blue triangles in the top panels of Figure 2.
The green long-dashed line in Figure 2(a) denotes the bar detection limit ($R_{\rm lim}$) where we strictly define as FWHM$/<Bar_{\rm min}/Bar_{\rm maj}>$,
%\begin{equation}
%R_{lim} = {{FWHM} \over {\left< \frac{Bar_{maj}} {Bar_{min}} \right>}},
%\end{equation}
where $<Bar_{\rm min}/Bar_{\rm maj}>$ is the mean of the bar minor to major axis ratio. We adopt $<Bar_{\rm min}/Bar_{\rm maj}>$ as 0.35 using the median bar ellipticity from the literature (\citealt{gadotti_11, kim_14}). The FWHM of the COSMOS F814W images is $0.09''$. Thus, $R_{\rm lim}$ is set to $0.26''$.

Figure 2(b) shows the redshift versus deprojected bar length in kpc. We use the $\rm R_{\rm bar}$ as deprojected bar length throughout this paper. Bar length is deprojected analytically. We employed the deprojection method developed by \citet{gadotti_07}. In this method, the key parameters to calculate the deprojected bar length are the inclination of the galaxy, the ellipticity of the bar and the angle between the bar and the line of nodes. We use the results from the GALFIT to calculate the deprojected bar length.

In order to see if there is any trend, we bin the data using redshifts with the step of $0.1$ from $z=0.2$ to $z=0.7$ and the last bin includes galaxies from $0.7 \leq z <0.835$. The mean bar length at each redshift bin is plotted with a large red open circle and blue filled triangle for massive and less massive galaxies, respectively. The orange and light blue shaded regions represent the standard deviation at each redshift bin centered at the mean bar length. Overlapped regions appear in green. The number of galaxies at each bin is $N=[11,41,29,31,51,26]$ for massive galaxies and $N=[31,65,31,22,31,10]$ for less massive galaxies.

%% CHANGE HERE!!!  %%%%%%%
%For massive galaxies, the absolute bar length shows a declining trend with redshift in Figure 2(b). The decrease is mainly driven at $0.2 < z \leq 0.4$ and $0.6 < z \leq 0.835$ by both ends of the redshift bins, while in intermediate bins the bar length remains rather constant.\textbf{Although the scatter is large, we find that there is a slight increase of the bar length in the lowest redshift bin ($0.2< z \leq 0.3$) for massive galaxies.}
%%%%%%%%%%%%%%%%%%%%%%%

For massive galaxies, it seems that there is a slight increase of the bar length in the lowest redshift bin ($0.2< z \leq 0.3$). 
In that lowest bin, however, there are only 11 massive galaxies and the standard deviation is too large to draw a firm conclusion, and thus the slight increase is statistically insignificant.
To show this, we bin the data differently by taking equal number of galaxies in each bin and find that there is no change in bar length with redshift at all redshift bins.
Also for less massive galaxies, no clear trend is found with redshift. 
We note that an earlier study by \citet{perez_12} reached a similar conclusion with 44 barred galaxies with an outer ring. They find that $R_{\rm CR}, R_{\rm bar}$ and their ratio $\cal R$ show similar distribution at $0<z<0.8$, and do not show any clear trend.
Thus, our results are in line with \citet{perez_12}.

We also plot the bar length versus the galaxy mass in the bottom panels of Figure 2. Galaxies are color-coded by T-type in Figure 2(c) and redshift in Figure 2(d). Linear fit to the data is plotted in black long dashed line. Using ordinary least square bisector fit, we find that galaxies in our sample follow $log(R_{\rm bar})= 0.74 \times log (M_{\ast}/M_{\odot}) -7.11$.

%The gray dotted lines in Figure 2(c) and (d) depict the fitted bar length as a function of the galaxy mass from \citet{erwin_19}, which used the data from the Spitzer Survey of the Stellar Structure of Galaxies (\s4g, \citealt{sheth_10, herrera_endoqui_15}). 

For a comparison with nearby galaxies, we took the bar length -- galaxy mass fit from \citet{erwin_19}, which used the data from the Spitzer Survey of the Stellar Structure of Galaxies (\s4g, \citealt{sheth_10, herrera_endoqui_15}), and plot them with gray dotted lines in Figure 2(c) and (d). 
Compared to the results from the nearby galaxies (\citealt{erwin_19}), the bar lengths of our samples are slightly larger at each mass bin and show an upward shift. 
Note that the bar lengths used in \citet{erwin_19} are visual measurements from \citet{herrera_endoqui_15} which are known to correlate well with the bar lengths from maximum ellipticity (\citealt{herrera_endoqui_15, diaz_garcia_16}) and normally shorter than the bar length measurements we adopted (\citealt{wozniak_95, athanassoula_02a, laurikainen_02_2mass, erwin_05_bar}). 
As we already have ellipticity profiles from our analysis, we have extracted the bar lengths from maximum ellipticity and compared them with those of \citet{erwin_19}. We find that the offset diminishes while there is a slight difference in the slope. The ordinary least square bisector fit for bar length from maximum ellipticity and the galaxy mass is $log(R_{\rm bar})= 1.06 \times log (M_{\ast}/M_{\odot}) - 10.67$.
Figure 2(c) shows a slight difference between galaxy types: early-type disk galaxies tend to host shorter bars at each bin. However, no such clear discrepancy with redshift is found in Figure 2(d). We present the ordinary least square bisector fit results for each galaxy type and redshift group in Table 1. 

It is necessary to take into account the effect of the change in resolution and wavelength in bar length. As we have estimated in Section 3.1 and 3.2, the resolution effects amounts to $1.5 \sim 4.3\%$ on average, and the wavelength effect is less than $5\%$ on average. We plot the representative error bars of $5\%$ at the mean bar length in Figure 2.

%%%%%%%%%%%%%%%%%%%%%%%%%%%%%%%%%%%%%%%%
\begin{deluxetable}{cccccc}

\tabletypesize{\small}
\tablecaption{Coefficients of the fit from bar radius and galaxy mass}
\tablenum{1}
\tablehead{\colhead{$\rm R_{\rm bar}$} & \colhead{Group} & \colhead{$C_0$} & \colhead{$C_0$ err} & \colhead{$C_1$} & \colhead{$C_1$ err} }
\startdata
          &   All                          &  -7.11   &  0.32  &  0.74  &  0.03 \\
Our  &  $T \leq 2$                  &  -8.48  &  0.42  &  0.85  &  0.04  \\
measurements   &  $T > 2 $   &  -6.88  &  0.41  &  0.72  &  0.04  \\
     &  $0.2 \leq z < 0.5$        &  -7.13  &  0.40  &  0.74 & 0.04  \\
     &  $0.5 \leq z < 0.835$  &  -7.58  &  0.50  &  0.77 & 0.05  \\
\hline \\
                  &  All               &  -10.67  &  0.71  &  1.06  &  0.07 \\
Max            &   $T \leq 2$   &  -12.26  &  1.62  &  1.20  &  0.15  \\
ellipticity     &   $T > 2 $     &  -7.58   &  0.39  &  0.78  &  0.04  \\
   &   $0.2 \leq z < 0.5$      &  -10.13  &  1.08  &  1.02  &  0.10  \\
   &   $0.5 \leq z < 0.835$  &  -11.67  &  1.09  &  1.15  &  0.10 \\

\enddata
\tablecomments{The coefficients of the bar radius and galaxy mass using ordinary least square bisector fit. The coefficients are of $log(R_{\rm bar})= \rm C_1 \times log (M_{\ast}/M_{\odot}) + \rm C_0$.}
\end{deluxetable}
%%%%%%%%%%%%%%%%%%%%%%%%%%%%%%%%%%%%%%%%

%% Fig 3 %%%%%%%%%%%%%%%%%%%%%%%%%%%%%%
Figure 3 presents various galaxy characteristic radii, such as disk scale length ($h$), bulge effective radius ($R_{\rm eff,\rm bul}$), and the radius containing $50\%$ and $90\%$ of the total galaxy luminosity ($R_{50}$ and $R_{90}$, respectively). We take $h$ and $R_{\rm eff,\rm bul}$ from the GALFIT decompositions. We estimate $R_{50}$ and $R_{90}$ by using results from the ellipse fittings. The lengths $h$ and $R_{50}$ range from 2 to 13 kpc, and $R_{90}$ spans from 5 to 25 kpc. As our sample contains bulgeless galaxies, $R_{\rm eff,\rm bul}$ ranges from 0 to 2 kpc.
%Our data sample do not show any cosmic evolution of characteristic radius of galaxy. 

%% Fig 4 %%%%%%%%%%%%%%%%%%%%%%%%%%%%%%
We explore which characteristic radius of a galaxy is most strongly correlated with the bar length. We plot $R_{\rm bar}$ versus $h$, $R_{\rm eff,\rm bul}$, $R_{50}$, and $R_{90}$ in Figure 4. The sample galaxies are divided into two groups of redshift: Galaxies at $0.2 < z \leq 0.5$ are plotted in red circles and those at $0.5 < z \leq 0.835$ are plotted in blue triangles.
We also present the Spearman coefficient ($\rho$) and significance ($P$) in each panel. The $\rho$ and $P$ of the two redshift groups are quite similar, implying that the bar and the disk size correlations are almost the same for the two redshift groups.
$R_{\rm bar}$ shows fairly good correlations with the characteristic lengths ($R_{\rm eff,\rm bul}$, $h$, $R_{50}$, $R_{90}$) with stronger correlations with disk-related parameters, $h$ and $R_{90}$. 
Figure 4 shows that the bar--disk size correlation is already in place at $0.5 < z \leq 0.835$, and the tightness of the relation is not much different from that at $0.2 < z \leq 0.5$.
The linear fit between each characteristic radius and $R_{\rm bar}$ for the whole sample is noted in the botton right of each panel in Figure 4.

Studies on nearby galaxies commonly employ $R_{25}$ and $R_{25.5}$, which are the radii where the surface brightness of the galaxy reach 25 and 25.5 $\rm mag/\rm arcsec^2$, respectively. Hence, we have attempted to estimate $R_{25}$ and $R_{25.5}$ for easy comparison with the studies on nearby galaxies. However, due to the cosmological surface brightness dimming, the surface brightness decreases by $(1+z)^{-4}$, amounting to $\sim$2 to 10 times from z= 0.2 to 0.835. 
It is possible to correct the surface brightness dimming, but to estimate $R_{25}$ and $R_{25.5}$ we need to make a further assumption, namely to assume the galaxy light profile.
Moreover, some galaxies at $z>0.7$ are barely observable at the surface brightness level of $25$ or $25.5 \rm mag/\rm arcsec^2$. Thus, we opt not to utilize $R_{25}$ and $R_{25.5}$ in our study.

%% Fig 5 %%%%%%%%%%%%%%%%%%%%%%%%%%%%%%
We now explore the cosmic evolution of the bar lengths normalized with $h$, $R_{\rm eff,\rm bul}$, $R_{50}$, and $R_{90}$ as a function of the redshift and plot them in Figure 5. The normalized bar length remains constant over time and does not show any clear redshift evolution. We divide the sample by galaxy mass to examine the galaxy mass dependence of the cosmic evolution. Massive galaxies are plotted in orange open circles and less massives ones in blue filled triangles. The black squares represent the mean value of all galaxies at each redshift bin. The gray shades cover the standard deviation at each bin. We present the Spearman's rank correlation coefficients and the significance in the upper right in each panel. Overall, we do not find clear evidence for cosmic evolution. 

%Fig5(b): 
Interestingly, there is a slight yet systematic decrease of the mean $R_{\rm bar}/R_{\rm eff,\rm bul}$ with redshift. 
We calculate the Spearman's correlation coefficient between the mean $R_{\rm bar}/R_{\rm eff,\rm bul}$ and z to find that $\rho= - 0.94$ and $P = 0.0048$. This indicates that the mean value of $R_{\rm bar}/R_{\rm eff,\rm bul}$ shows a statistically significant decrease with redshift. 

To compare with studies of nearby galaxies, we retrieve $R_{\rm bar}$ and galaxy characteristic radius from \citealt{erwin_05_bar} ($D< 30$ Mpc, purple squares) and the SDSS (\citealt{gadotti_09}, $z<0.007$, green squares) and the \s4g (\citealt{kim_14}, blue square, \citealt{salo_15}, red square, \citealt{diaz_garcia_16}, light blue squares, $D < 40$ Mpc), and calculate the median of the normalized bar length and plot them in Figure 5. The $R_{\rm bar}/h$ of galaxies at $0.2 < z \leq 0.835$ and nearby agree well. 
\citet{aguerri_05} obtained the mean $<R_{\rm bar}/h>  = 1.21 \pm 0.08$ for 14 nearby SB0 galaxies, when they included a lens component in their surface bright fitting analysis. However, $<R_{\rm bar}/h>$  increases to $1.42 \pm 0.09$ when they excluded the lens component from their decompositions. 
We do not adopt a lens component in our 2D decompositions. Thus, our $<R_{\rm bar}/h>$ agree fairly well with \citet{aguerri_05}.   
\citet{erwin_05_bar} find that $R_{\rm bar}/h$ ranges from 0.2 to 2.5, and the mean $<R_{\rm bar}/h>$ changes with the Hubble type, i.e., higher in early type barred galaxies ($<R_{\rm bar}/h> = 1.4 $) and lower in late type barred galaxies ($<R_{\rm bar}/h> = 0.6$). The drop in $<R_{\rm bar}/h>$ along the Hubble type is remarkable for galaxies with $T \ge 4$ (\citealt{diaz_garcia_16}). Although we include late-type disk galaxies, galaxies with $T \ge 4$ are rare in our sample. Thus, $<R_{\rm bar}/h>$ of nearby and $0.2 < z \leq 0.835$ galaxies are in agreement. 

While the mean $R_{\rm bar}/R_{\rm eff,\rm bul}$ from \citet{gadotti_09} and \citet{kim_14} are slightly lower than that from $0.2 < z < 0.835$, they are still within the range of the $R_{\rm bar}/R_{\rm eff,\rm bul}$ distribution. 
The normalized bar length from \citet{algorry_17}, which used the EAGLE cosmological hydrodynamical simulation, is plotted in the Figures 5(c) and (d). Although the median $R_{\rm bar}/R_{50}$ of the EAGLE study is slightly higher than that at each redshift, the EAGLE predictions are still within the range of our results. This also applies to $R_{\rm bar}/R_{90}$ except that the median $R_{\rm bar}/R_{90}$ is slightly smaller than that in our results.

% Why Rbar/Reff,bul discrepance?
%kim_14 ( 9 < log Mass < 11),  gadotti_09 (10 < log Mass < 11.5), Salo_15 (9.5 < logmass < 11.5) -> not by mass

We note that there are outliers in the normalized bar length in Figure 5. In particular, some galaxies have remarkably high $R_{\rm bar}/R_{\rm eff,\rm bul}$. Our visual inspections showed that the bulges of those galaxies are considerably smaller than average;
thus, small $R_{\rm eff,\rm bul}$ values led to high $R_{\rm bar}/R_{\rm eff,\rm bul}$. Moreover, a few galaxies have exceptionally high $R_{\rm bar}/R_{50}$. A visual check allowed us to find that these bars are unusually brighter than the underlying disk. It almost looks as if no significant light comes from the disk. 
Such galaxies were first noted in observations by \citet{gadotti_03}, and their formation in simulations were discussed in A03, as due to very strong angular momentum redistribution in the galaxy. The morphology of the outliers of $R_{\rm bar}/R_{90}$ also resembles those of the outliers in $R_{\rm bar}/R_{50}$. The origin of such unusual barred galaxies with a faint disk is not yet well known. Using the IllustriTNG simulations (\citealt{pillepich_18, springel_18, nelson_19}), \citet{lokas_20c, lokas_20b} recently reported that ``elongated bar-like galaxies without a significant disk component'' are often formed through tidal interactions with a galaxy cluster or they might be infalling to the cluster for the first time. The properties of such galaxies well deserve a further study, but are beyond the scope of this paper. We will explore them in the future. 

\subsection{Cosmic Evolution of the Light Deficit in the Star Formation Desert Region}
\label{sec:res2}
%% Fig 6 %%%%%%%%%%%%%%%%%%%%%%%%%%%%%%
Numerical simulations have shown that as a galaxy evolves, its bar grows longer and stronger with time (e.g., \citealt{combes_93, athanassoula_02a}; A13 for a review). In particular, bars capture disk stars in the vicinity of the bar within the bar radius to build longer and stronger bars (A03). As a result of this capture, the light deficit around the bar becomes more prominent. This can be an evidence of a bar-driven secular evolution. The light deficit is a strong function of the bar length and strength among nearby galaxies (\citealt{kim_16, buta_17b}). In this section, we examine how the light deficit of the star formation desert evolves with time. As defined in \citet{kim_16}, we calculate $\rm Max(\Delta \mu)$, the maximum difference between the surface brightness along the bar major and minor axes, as follow: 

\begin{equation}
\rm Max(\Delta \mu) = 
\mid {\rm Max [\mu_{\rm bar \; \rm maj}(r) - \mu_{\rm bar \; \rm min}(r)]} \mid,  r<R_{\rm bar}.
\end{equation}

The surface brightness $\mu_{bar \; \rm maj}(r)$ is that along the bar major axis while $\mu_{bar \; \rm min}(r)$ is measured along the minor axis. We plot $\rm Max(\Delta \mu$) as a function of z, bar length, and bar-to-total luminosity ratio (Bar/T) in Figure 6.  $\rm Max(\Delta \mu$) do not show any clear evolution at $0.2 < z \leq 0.835$.  
%However, massive galaxies exhibit a slight decrease of $\rm Max(\Delta \mu$) with z ( Spearman's rank correlation coefficient $\rho=-0.2$ and $P<0.003$). The lower boundary of the distribution shows a slight decrease with z such that a lack of low $\rm Max(\Delta \mu$) at low z is observed. However, the scatter is too large to give any strong statement on the cosmic evolution.

$\rm Max(\Delta \mu$) increases with the normalized bar length ($R_{\rm bar}/h$) and Bar/T in Figures 6(b) and (c), albeit with some outliers. The Spearman correlation coefficients ($\rho$) and the statistical significances ($P$) are presented at the top of each panel.
Galaxies with a longer bar and a higher Bar/T show a more prominent light deficit around the bar, implying that galaxies at $0.2 < z \leq 0.835$ already experience some degree of bar driven secular evolution. 

For nearby galaxies, both $\rm Max(\Delta \mu$)--($R_{\rm bar}/h$) and $\rm Max(\Delta \mu$)--Bar/T relations are tighter. The Spearman correlation coefficients ($\rho$) of both relations are roughly 0.7 for nearby galaxies (\citealt{kim_16}). However, galaxies at $0.2 < z \leq 0.835$ 
have hosted their bars during a 
relatively shorter period compared to nearby ones. Thus, it seems natural that both relations are less tight compared to those of the nearby ones.

Figure 6(c) shows a number of outliers with $\rm Max(\Delta \mu)>2$ (31 galaxies), standing out from the trend of Bar/T -- $\rm Max(\Delta \mu$). The majority of these outliers are $T>2$ (23 out of 31) and all of them are $T>1.5$, rather late-type barred galaxies.
One possible reason to explain this could be that it might be more difficult to capture and move disk stars into bar orbits for earlier-type galaxies due to stronger central spherical (i.e., bulge) components.

%\textcolor{blue}{Re-arrange the following later- here? or in discussion part?}
Our results are in line with the study on nearby galaxies that found higher $\rm Max (\Delta \mu$) in more evolved bars (\citealt{kim_16}). As galaxies evolve, bars become longer and stronger, and the light deficit around the bar becomes more prominent, i.e., $\rm Max(\Delta \mu$) increases. If we put this in a timeline, we can expect that $\rm Max(\Delta \mu$) decreases with lookback time (i.e., decrease with redshift). Therefore, if we track one galaxy along the redshift, or assume that the majority of the galaxies formed bars at a specific redshift, we can expect to see a clear downward trend of $\rm Max(\Delta \mu$) with z. However, the epoch of bar formation time in each galaxy differs from each other, and the amount of evolution (i.e., bar growth rate) varies depending on the galaxy properties. Thus, these would bring scatter in the evolutionary trend, which would complicate the pictures.

\subsection{Bar strength evolution}
\label{sec:res3}
%% Fig 7 %%%%%%%%%%%%%%%%%%%%%%%%%%%%%%
\citet{buta_17b} found that $\rm Max (\Delta \mu)$ is closely related to the maximum m = 2 Fourier relative intensity amplitude, $A_2$ for nearby galaxies. We check $\rm Max (\Delta \mu)$ versus $A_2$ in Figure 7 and confirm that they are tightly related for galaxies at $0.2 < z < 0.835$. $\rm Max (\Delta \mu)$ is also found to be closely related to another bar strength parameter, $Q_b$, but with some degree of scatter. We explore whether or not bar strengths ($A_2$ and $Q_b$) evolve with time in Figure 7(c) and (d). We find that $A_2$ and $Q_b$ remain rather constant over time. We divide the sample into two groups according to the galaxy mass. Yet, bar strengths remain constant in both mass bins.
\citet{perez_12} estimated the bar strength by using the maximum ellipticity for galaxies at $0<z<0.8$ and found that there is no clear evolution of bar strength with redshift. Our results are consistent with that of \citet{perez_12}.

\subsection{Bar length with Bulge/T and galaxy mass}
%% Fig 8 %%%%%%%%%%%%%%%%%%%%%%%%%%%%%%
Bar length is found to be associated with Bulge/T (\citealt{gadotti_11}) and bulge prominence (\citealt{hoyle_11}) for nearby galaxies. 
% With numerical simulations, it has been found that the size of the bulge and the bar length is closely related (\citealt{sellwood_80, athanassoula_80, athnassoula_03}). We also find that same results (not presented here)
It is also correlated with the galaxy mass (\citealt{diaz_garcia_16, erwin_19}) for nearby galaxies.
We examined how normalized bar lengths vary with Bulge/T and the galaxy mass for galaxies at $0.2 < z \leq 0.835$ in Figure 8. 
The normalized bar lengths remain constant along Bulge/T, except $R_{\rm bar}/R_{50}$ which shows a clear increase.
There is a caveat when we explore various radii with Bulge/T. In particular, $R_{50}$ is frequently used in the high-z galaxy studies (usually denoted as half light radius or effective radius), but, we find that $R_{50}$ itself is anti-correlated with Bulge/T. This is natural in a way that highly concentrated galaxies (high Bulge/T) should have smaller $R_{50}$. The Spearman's rank correlation coefficients of $R_{50}$ versus Bulge/T are $\rho = -0.44$ with $P< 10^{-5}$. We linearly fit $R_{50}$ and Bulge/T for galaxies with Bulge/T$>0$ and obtained a relation $-8.26 \times Bulge/T + 6.26$. We then simply normalize $R_{50}$ with the obtained linear relation, and plot them in the inset of Figure 8(b). The clear increase in $R_{\rm bar}/R_{50}$ with Bulge/T flattens out after the normalization. This implies that the increase of $R_{\rm bar}/R_{50}$ with Bulge/T is mainly driven by the anti-correlation of $R_{50}$ and Bulge/T. 
This is a natural result that a more strongly concentrated galaxy should have a smaller $R_{50}$. Therefore, we should be careful in using $R_{50}$, widely used in exploring the size of galaxies for high-z studies, as a normalizer when we examine galaxy properties related to Bulge/T. 
We divide the sample into two groups of z to examine whether a difference exists in the trend with the redshift.
Only small differences have been found between the two redshift groups, which are still within the standard deviation of each Bulge/T bin. Thus, this ensures the absence of a clear redshift evolution in the normalized bar lengths versus Bulge/T.

Next, we examine the normalized bar length versus galaxy mass in the bottom panels of Figure 8. All normalized bar lengths stay at a constant level from $\log(M/M_*)=10$ to $11.4$, again showing no redshift evolution in the normalized bar lengths versus galaxy mass.

%\label{sec:res3}                                                                        

\section{Discussion}

Bar formation and evolution is a complex subject and it is not easy to
develop this fully here.
We will only discuss some specific aspects, 
which are necessary for us to compare simulations to our results on
the evolution of bar length and strength, as obtained from galaxies of
the COSMOS survey. We find that, on average over all the sample,
these quantities do not increase  with time, a result that may, at first sight, seem to clash with 
numerous simulation results. For a fuller view of bar evolution,
the reader can consult the review of A13.  

\subsection{How can the bar lose angular momentum? }
The bar can loose its angular momentum in many ways, of which three main ones have
been discussed in A03. We will here give a simplified description of
these processes.

1. The angular velocity of the bar, usually referred to as its
pattern speed, can decrease, i.e. the bar will slow down. This has been witnessed in 
most N-body simulations and has been well studied. Unfortunately, it
can not be easily obtained from observations, where, at the best, we
measure the pattern speed at only one given time.  

2. The bar can trap material from its close neighborhood in phase space. Since both the periodic and the regular orbits beyond the bar are, in general, less elongated than those within the bar, the whole process will be accompanied by a loss of angular momentum. 
In observations, this should be witnessed by a lengthening of the bar with increasing evolution time\footnote{Evolution time is the time since bar formation and increases with decreasing redshift} and of course it can not be directly observed. 
Some information, however, can be obtained by comparing galaxies at different stages of their evolution, and this is the path we are exploring here. Alternatively, the trapping is not done in the region somewhat beyond the bar, but in the region somewhat within it. In this case it adds to the population of orbits within the bar and would change the density distribution and the bar strength, but not necessarily the length of the bar.

3. The bar orbits can become yet more elongated, i.e. a given elongated
bar orbit will become thinner. Up to date, this has not been much studied, neither  observationally, nor using simulations. In the latter, one can
get such an impression from short video clips of the bar evolution (Athanassoula et al. in prep).
But no study has been made and, therefore, it is not clear whether
this impression comes from the bar becoming thinner, or by its
becoming longer, or both together. 

%As described above, the bar can lose its angular momentum in three different ways and,
To our knowledge, there is no in depth study about which of the three is
preferentially followed and what properties of the galaxy determine that, 
although some useful information can be found in A03, \citet{athanassoula_13} and \citet{athanassoula_14}, whose main aim, however, was to cover specific aspects of the problem.

We should, nevertheless, keep in mind that, in order to have secular evolution,
it is not necessary that all three of the above are strongly present.
Thus, showing that one of the three shows hardly any evolution is
not sufficient to establish that there is no evolution.

%=========================

\subsection{What determines the amount of angular momentum exchanged
within the galaxy?} 

Using a large number of N-body simulations, it has been shown (A03; A13; \citealt{athanassoula_13}) that there is a strong relation between the bar strength and 
 the angular momentum redistribution within the galaxy.
The amount of redistribution depends on how much angular
momentum can be emitted from the inner region (bar/bulge region) of the
galaxy and how much can be absorbed in the outer regions (partly by the outer
disk, but mainly by the halo). Furthermore, the quantities emitted or absorbed
depend on the density and velocity distribution of baryonic and of dark  
matter in these regions.
%, be it baryonic, or dark matter.   

Given all the relevant information on the phase space distribution, 
it is possible to calculate the amount of angular momentum redistribution
within a model galaxy (\citealt{lyndenbell_72, tremaine_84, weinberg_85}; A03)
The opposite, however, is not 
true, so that given the strength of the bar,
one can not obtain information on any single property of the model or
the galaxy (A03; \citealt{athanassoula_14}). 
%So, a given bar strength can be obtained with
%more than one mix of the stellar, gaseous and dark matter mass
%distribution of the galactic components, stars gas and dark matter, sh     
Indeed, we can achieve a given amount of bar growth -- i.e. a given bar strength as a function of time -- in more than one way, since we are in a
multi-dimensional parameter space. 

A number of properties of the galaxy affect its angular momentum redistribution and therefore its evolution, for example the  
spin of the dark matter halo (\citealt{long_14, collier_18}, see also \citealt{saha_13}). Bars experience less growth in length and strength for a higher halo spin,
as spinning halo leads to a decreased transfer of angular momentum between the disc and its parent halo (\citealt{long_14, collier_18}). Likewise, the shape of the halo plays a role in the bar evolution (\citealt{machado_atha_10, athanassoula_13, collier_18}). The disk thickness also affects the bar strength (\citealt{klypin_09}).

In order to achieve a very low bar growth
(which is of particular interest to us here) we can, for example,
minimize the amount of angular momentum that the halo can absorb
\citep{athanassoula_02b, athanassoula_02a}, 
and/or include a gaseous galactic disc component \citep{villa_vargas_10, athanassoula_13, seo_19}, etc. 
Going through the literature one can find a number of cases where 
such a low growth rate was achieved,
both in cases with no gas, as for example in \citet{athanassoula_02c, machado_atha_10},
and cases with it, 
e.g. \citep{berentzen_07, seo_19}.

The most convincing cases, however, can be found in figures 7 and 8
of \citet{athanassoula_13}. This work studied two effects, that of 
a gaseous disk and that of halo triaxiality, using a 2D grid of fifteen models with five
different initial gas fractions and three different initial halo triaxialities. 
In order to be able to make comparisons, all remaining properties and parameters were kept constant. Thus, any differences between models would be due only to the effect of the gas, or to that of the halo shape (see \citet{athanassoula_13} for more detailed information).
In six out of the fifteen cases, the bar strength during the secular
evolution time can be considered as
non-evolving. This showed that, by varying the amount, or the distribution of gas present, or the halo triaxiality, or both of these two properties, it is possible to have very little  secular evolution, or practically none,
in good agreement with what we found here with the COSMOS data. 
Using different values for some other parameters of the model -- such as disk thickness or different radial distributions for the dark matter density or Toomre's parameter Q -- we could hinder the bar growth further and find more cases compatible with no bar growth.

These simulations also show that the strongest bar growth occurs for cases with initially
spherical halos and no gas component, both of which are unlikely at
higher redshifts. Indeed, observations indicate that galaxies are more
gas-rich with increasing redshift \citep{genzel_15}. Furthermore, close
interactions and mergers make galactic halos triaxial (e.g. \citealt{mcmillan_07}). Thus, of 
the initial conditions of the \citet{athanassoula_13} simulations, those that are more compatible with what we know about galaxies at the high end of redshift range considered here,
are those which have the least bar growth. So our results in this paper are not in
contradiction with numerical simulations, provided these include an adequate number of the principal factors determining the bar strength. 

We can thus conclude that there is at present no clear qualitative disagreement
between our results and those of relevant N-body simulations,
as well as with previous observational studies.  
Nevertheless, further work is necessary,
both from the simulation and the observation side, to reach better
understanding of bar evolution and its effect on the galaxy evolution.
In particular, it would be useful to extend this comparison to the normalized bar length. Given, however, the difficulties with bar length measurements (Section 3.1) and the insufficient measurements available from simulations for the bar length and, particularly, for the disc scale length (see, however, \citealt{debattista_06}), we must defer this to a future study.

%We note that an earlier study by \citet{perez_12} reached a similar conclusion, using a different approach, and a rather small sample of 18 galaxies from the SDSS and 26 galaxies from the COSMOS survey, selected to be nearly face-on. They found that for 44 barred galaxies with an outer ring, $R_{\rm CR}, R_{\rm bar}$ and their ratio $\cal R$ do not change considerably at $0<z<0.8$. They argue that if bars are long-lasting, their results imply that no substantial angular momentum exchange between the bar and dark matter halo has occurred. Taken at face value, ours results are in line with \citet{perez_12}.

\subsection{Comparing observations to simulation results}
%\subsection{Comparing simulation results to observations}
Our results show that the mean bar length and strength at each redshift bin do not evolve with redshift at $0.2 < z < 0.835$. This is consistent with the result of \citet{perez_12} who found that the bar length, strength and also the bar pattern speed have not evolved over the last 7 Gyrs.

The time needed for a bar to start forming and then to reach the
secular evolution phase depends on many parameters and properties of
the host galaxy and of its environment (e.g., A03; \citealt{berentzen_07}; A13; \citealt{renaud_13, seo_19}).
Hence, in any redshift bin there are bars at different stages of their evolution and the mean
values in that bin will give information on the evolution of
individual bars convolved by the distribution over bar ages/stages. 

In simulations, to measure the evolution of a given bar between
times $t_1$ and $t_2$, one simply measures the change in bar length or
strength during the time $dt = t_2 - t_1$. This is not the case for
observations, where it is necessary to take averages over all
galaxies, i.e. to take into account age and evolutionary stage. 
The resulting average bar parameters then includes a number of bars 
with a considerably shorter dt, so that one obtains a lower value for 
the evolution than from a single simulation.

One should also take into account that in realistic simulations
(e.g. including star-forming gas, and/or haloes with appropriate shape
and kinematics) the secular evolution of the bar proceeds at a relatively
low pace. Moreover, in the above, we have not taken into account the fact
that barred galaxies, like all other galaxies, will undergo mergings
which could destroy or at least strongly perturb the bar (\citealt{Pfenniger_91, athanassoula_99, berentzen_03, ghosh_21}) and that these are more important at
higher redshifts.

Thus, by taking into consideration that in observations we take
averages in bins with a given redshift range (which is the only
possibility with the available data), that galaxies from realistic
simulations show relatively little evolution and that mergings are more
common at higher redshifts, we conclude that these weighted averages
will be considerably smaller than the corresponding values for
individual galaxies and may well not show large differences of bar
strength and length as a function of redshift, even though bars in
individual galaxies do. So our results do not necessarily show that
there is no evolution in individual galaxies, and, likewise, they do
not lead to the conclusion that the no-evolution possibility is
wrong.

\subsection{Little cosmic evolution in cosmological simulations}
There have been several recent studies which explored the evolution of bar length and strength based on cosmological simulations, and we briefly discuss  
them here.

\citet{okamoto_15} ran cosmological hydrodynamical zoom re-simulations for two Milky way-mass galaxies and find that `overall' bar lengths increase with time from 0 to $3 \sim 6 $ kpc from $z\sim 1.5$ to $z\sim 0$. The simulated galaxy with a higher central mass concentration, which resembles early-type disk galaxies, shows a remarkable increase in bar length, but the one with a lower concentration shows only a mild evolution in the bar length.
This is in good agreement with earlier results from simulations with idealized initial conditions (\citealt{athanassoula_02a}) and was explained by Okamoto et al. as due to the amount of material at the resonances.  This regulates the angular momentum exchange, as initially found and discussed in A03.

Using IllustrisTNG simulations (e.g., \citealt{springel_18}), \citet{rosas_guevara_20} analyzed 270 disk galaxies, and found that $40\%$ are barred. They show a strong bar case where the bar becomes stronger with time, but hardly increases in extent. They find from their statistics that the age of strong bars increases with increasing stellar mass, although with a large scatter, and that weak bars formed only recently. 
\citet{zhao_20} examined approximately 4000 barred galaxies from the TNG100 simulation and found that the bar sizes have grown by 0.17 dex from $\sim$2 kpc at $z = 1$ to $\sim$3 kpc at $z = 0$. They compared their analysis with the COSMOS sample in their Figure 8, and find a disagreement in the bar fraction as function of time. They also found that galaxies have maintained a virtually constant bar ellipticity of $\epsilon_{\rm max} = 0.5$ from $z = 1$ to $z = 0$.

\citet{fragkoudi_21} analyzed barred galaxies from the Auriga cosmological simulations (\citealt{grand_17}) which successfully reproduced realistic bar features at $z=0$ (\citealt{blazquez-calero_20}) as well as the observations of the fraction of barred galaxies as a function of redshift (\citealt{fragkoudi_20}). 
In the sample of five model galaxies from the Auriga sample, only one galaxy can be considered as having a global increase of the bar strength with time (See Figure B.1 of \citealt{fragkoudi_21}), while two galaxies have a bar strength compatible
with no such increase and for the remaining two no safe conclusion can be
reached.

\subsection{Necessity for a wide range of redshift and mass?}
%}
We have explored galaxies only up to $z \sim 0.835$ due to the bandshift effect because the wavelength of F814W (8140 \AA) becomes almost the B-band (4436 \AA) at z = 0.835. Exploring the bar structures using our dataset bluer than the B-band is not possible. 
%Once bars form, they bring the gas inwards to the central region of galaxies. Hence, one can assume that the stars of the nuclear disk and nuclear bar start to form only after the large scale bar is formed (\citealt{gadotti_15, neumann_20}). While many galaxies are found to host young nuclear components (e.g., \citealt{delorenzo_caceres_19, bittner_20}), a few galaxies are found to host old ($\sim$ 10 Gyrs) nuclear components which suggest the bar formation epoch as $z \sim 1.8$ (\citealt{gadotti_15, bittner_20}). Therefore, there might be the major evolution for those galaxies even at z$>$0.835. 
A further study using the CANDELS (\citealt{grogin_11, koekemoer_11}) with F160W or the upcoming COSMOS-Webb Survey using the James Webb Space Telescope (JWST) will facilitate our expansion of the redshift range over z$\sim$ 0.835.

For a consistent sample selection to cover our redshift range, we are limited to cover galaxies with the mass range of $\log(M/M_{\ast}) = 10 \sim 11.4$. 
More massive galaxies are found to acquire their bars at higher redshifts, implying the downsizing of the bar formation (\citealt{sheth_08}).
Thus, a more massive galaxy might have had more time to evolve.
% already have passed the fast growth evolutionary phase, and may already have reached its later evolutionary phase where the amount of the bar growth is insignificant.
Therefore, a future study on a very low- and high- mass galaxies will give us useful information in understanding the cosmic evolution of barred galaxies.

\section{Summary and Conclusions}

We have examined the cosmic evolution of 379 barred galaxies at $0.2 < z \leq 0.835$ with  $10.0 \leq \log (M_{\ast}/M_{\odot}) \leq 11.4$  using F814W images from the COSMOS survey. 
Our main results are summarized below:

\begin{enumerate}
\item Bar lengths and strengths meaned over redshift bins do not show any clear trend with redshift, but remain roughly constant for the whole redshift range we explored.

\item The normalized bar lengths ($R_{\rm bar}/h, R_{\rm bar}/R_{\rm 50}$ and $R_{\rm bar}/R_{\rm 90}$) also remain constant and independent of the redshift, not showing any trend. $R_{\rm bar}/h$ and $R_{\rm bar}/R_{\rm 90}$ remain constant and independent of the Bulge/T and galaxy mass. This implies that as galaxies evolve, bars grow in proportion to the disk size.

\item
The bar lengths are strongly related to the galaxy mass and disk scale lengths, $R_{50}$ and $R_{90}$. Bars and disks are already closely linked at $0.5 < z \leq 0.835$ where not much time is expected to have passed since the bar formation. This could imply that bars inherit their size properties from disks. However, bars and disks are also tightly linked at $0.2 < z \leq 0.5$. Hence, the coupling between the bar and the disk should be prompt and efficient. 

\item
We review results of numerical simulations and argue that there is no clear disagreement between them and the results we found here. 

\end{enumerate}

%-----------------------------------------------------------------------------------------------------

\acknowledgments

We are grateful to the referee for the constructive comments based on a thorough review. 
TK was supported by the Basic Science Research Program through the National Research Foundation of Korea (NRF) funded by the Ministry of Education (NRF-2019R1A6A3A01092024). 
EA and AB acknowledge financial support from the CNES (Centre National d$'$Etudes Spatiales France). MGP and YHL acknowledge support from Basic Science Research Program through the National Research Foundation of Korea (NRF) funded by the Ministry of Education (NRF-2019R1I1A3A02062242). We thank the BK21 Plus of National Research Foundation of Korea (22A20130000179).  
This research is based on observations made with the NASA/ESA Hubble Space Telescope obtained from the Space Telescope Science Institute, which is operated by the Association of Universities for Research in Astronomy, Inc., under NASA contract NAS 5–26555. 

{\it Facilities:} \facility{ The {\em{Hubble}} Space Telescope}

\bibliography{tkim_bar}
%\bibliography{abref}
%\addbibresource{addrefs.bib}

\clearpage
\end{document}